\pdfoutput=1
\documentclass[aps,prd,amsmath,amsfonts,amssymb,eqsecnum,nofootinbib,showpacs,
  floatfix,secnumarabic,preprintnumbers,superscriptaddress,nobalancelastpage,
  onecolumn,notitlepage]{revtex4-1}
\usepackage{color,graphicx,hyperref,slashed}
\usepackage[utf8]{inputenc}
\hypersetup{
   bookmarks=true,         % show bookmarks bar?
   unicode=true,          % non-Latin characters in Acrobat�s bookmarks
   pdftoolbar=true,        % show Acrobat�s toolbar?
   pdfmenubar=true,        % show Acrobat�s menu?
   pdffitwindow=false,     % window fit to page when opened
   pdfstartview={FitH},    % fits the width of the page to the window
   pdftitle={My title},    % title
   pdfauthor={Author},     % author
   pdfsubject={Subject},   % subject of the document
   pdfcreator={Creator},   % creator of the document
   pdfproducer={Producer}, % producer of the document
   pdfkeywords={keyword1} {key2} {key3}, % list of keywords
   pdfnewwindow=true,      % links in new PDF window
   colorlinks=true,       % false: boxed links; true: colored links
   linkcolor=blue,          % color of internal links (change box color with linkbordercolor)
   citecolor=magenta,        % color of links to bibliography
   filecolor=magenta,      % color of file links
   urlcolor=cyan           % color of external links
}

\newcommand{\Tr}[1]{\text{Tr}\big[#1\big]}
\def\be{\begin{equation}}
\def\ee{\end{equation}}
\def\bsp#1\esp{\begin{split}#1\end{split}}
\newcommand{\met}{{\slashed{E}_T}}
\def\bpm{\begin{pmatrix}}
\def\epm{\end{pmatrix}}
\def\ie{{\it i.e.}}
\definecolor{darkred}{rgb}{0.7, 0.0, 0.0}

\begin{document}
\date{\today}
%\preprint{IP/BBSR/2015-4}

\title{Investigating the scalar sector of left-right symmetric models
  with leptonic probes}

\author{Debasish Borah}
\email{dborah@iitg.ac.in}
\affiliation{Department of Physics, Indian Institute of Technology Guwahati,
  Assam 781039, India}

\author{Benjamin Fuks}
\email{fuks@lpthe.jussieu.fr}
\affiliation{Sorbonne Universit\'e, CNRS, Laboratoire de Physique Th\'eorique et
  Hautes \'Energies, LPTHE, F-75005 Paris, France}
\affiliation{Institut Universitaire de France, 103 boulevard Saint-Michel,
  75005 Paris, France}

\author{Deepanjali Goswami}
\email{g.deepanjali@iitg.ac.in}
\affiliation{Department of Physics, Indian Institute of Technology Guwahati,
  Assam 781039, India}

\author{P. Poulose}
\email{poulose@iitg.ac.in}
\affiliation{Department of Physics, Indian Institute of Technology Guwahati,
  Assam 781039, India}

\begin{abstract}
We investigate the potential collider signatures of singly-charged and
doubly-charged Higgs bosons
such as those arising in minimal left-right symmetric models. Focusing on
multileptonic probes in the context of the high-luminosity run of the Large
Hadron Collider, we separately assess the advantages of the four-leptonic and
trileptonic final states for a representative benchmark setup designed by
considering a large set of experimental constraints. Our study
establishes possibilities of identifying singly-charged and doubly-charged
scalars at the Large Hadron Collider with a large significance, for luminosity
goals expected to be reached during the high-luminosity phase of the Large
Hadron Collider. We generalise our results and demonstrate that existing
limits can in principle be pushed much further in the heavy mass regime.
\end{abstract}

\maketitle

\section{Introduction}
The Standard Model of particle physics has been established as the most
successful theory describing the elementary particles and their interactions
(with the exception of gravity), especially after the discovery of a
Standard-Model-like Higgs boson at the Large Hadron Collider (LHC) in
2012~\cite{Aad:2012tfa,Chatrchyan:2012ufa}. Despite its successes, the Standard
Model features several conceptual issues and practical limitations that have
motivated the particle physics community to focus on theoretical developments
and experimental studies of physics beyond it. In particular, the
Standard Model cannot provide any explanation for the observed non-zero neutrino
masses and mixings~\cite{Agashe:2014kda}. In the Standard Model, the Higgs field
is responsible for the generation of the masses of all known fundamental
particles, but it cannot accommodate the tiny observed neutrino masses within
only renormalisable interactions. The situation nonetheless changes at the
non-renormalisable level since neutrino masses can be generated through the
dimension-five Weinberg operator~\cite{Weinberg:1979sa} that generally
arises, within a renormalisable ultraviolet-complete theory where new heavy
fields are introduced, through a seesaw mechanism. The different
realisations of such a mechanism can be broadly classified into three categories
named type~I~\cite{Minkowski:1977sc,GellMann:1980vs,Yanagida:1979as,%
Mohapatra:1979ia,Schechter:1980gr} (that relies only on right-handed neutrinos
coupling to the Higgs field), type~II~\cite{Mohapatra:1980yp,Lazarides:1980nt,%
Wetterich:1981bx,Schechter:1981cv} (that makes use of a new scalar field lying
in the adjoint representation of $SU(2)_L$) and type~III~\cite{Foot:1988aq}
(where at least two extra fermionic fields lying in the adjoint
representation of $SU(2)_L$ are included) seesaws.

In most common seesaw implementations, heavy fields are supplemented to the
Standard Model in an {\it ad-hoc} fashion so that the desired neutrino
properties are reproduced after the breaking of the electroweak symmetry.
Seesaw options where the
symmetries of the Standard Model are extended also exist, like in
left-right symmetric theories~\cite{Pati:1974yy,Mohapatra:1974hk,%
Mohapatra:1974gc,Senjanovic:1975rk,Mohapatra:1977mj,Senjanovic:1978ev,%
Mohapatra:1980qe,Lim:1981kv,Gunion:1989in,Deshpande:1990ip,FileviezPerez:2008sr,%}
Senjanovic:2016vxw}
where minimal and non-minimal realisations naturally feature type~I/II and
type~III seesaw
mechanisms respectively. In the left-right symmetric paradigm, the Abelian
hypercharge symmetry group of the Standard Model $U(1)_Y$ is extended to an
$SU(2)_R \times U(1)_{B-L}$ gauge group and the right-handed $SU(2)_L$ fermionic
singlets are collected into $SU(2)_R$ doublets,
which naturally requires the introduction of right-handed neutrino fields in the
spectrum. Focusing on minimal model building possibilities (and thus on type-I
and type-II seesaws), the extended gauge symmetry is spontaneously broken down
to the Standard Model gauge symmetry
thanks to a scalar field that lies in the adjoint representation of $SU(2)_R$ or
$SU(2)_L$ for a type~I and type~II seesaw mechanism respectively. This field
acquires a non-vanishing vacuum expectation value that yields
neutrino mass generation. In their most minimalistic form,
left-right symmetric theories are moreover symmetric under parity
transformations in the ultraviolet regime, although low-energy parity violation
arises after the spontaneous breaking of the left-right symmetry at a high
energy scale. Finally, such theories can also be embedded within an $SO(10)$
grand unified context and feature gauge coupling unification.

With a particle content exhibiting three extra vector bosons, three right-handed
neutrinos and several new scalar (Higgs) fields whose components possess either
a double, a single or a vanishing electric charge, minimal left-right symmetric
models (MLRSM) feature various collider signatures that can be used as probes
for new
physics~\cite{Pati:1974yy,Mohapatra:1974hk,Mohapatra:1974gc,Senjanovic:1975rk,%
Mohapatra:1977mj,Senjanovic:1978ev,Mohapatra:1980qe,Lim:1981kv,Gunion:1989in,%
Deshpande:1990ip,Senjanovic:2016vxw}.
The Higgs sector of the model has in particular been
recently investigated~\cite{Barenboim:2001vu,Polak:1991vf,%
Azuelos:2004mwa,Jung:2008pz,Bambhaniya:2013wza,Alloul:2013raa,Dutta:2014dba,
Bambhaniya:2014cia,Maiezza:2015lza,Bambhaniya:2015wna,Dev:2016dja,Babu:2016rcr,%
Maiezza:2016ybz}.
In the present work, we study several Higgs processes giving
rise to the production of four-lepton and three-lepton systems
and focus on setups where these final states are stemming
from the production and decay of either a pair of doubly-charged Higgs bosons,
or of an associated pair comprised of a doubly-charged Higgs boson and a
singly-charged Higgs or vector boson. Unlike most earlier studies, we consider
a framework where both type~I and type~II seesaw mechanisms are implemented and
contribute to neutrino mass generation. Equivalently, the neutral component of
the left-handed ($\Delta_L$) and right-handed ($\Delta_R$) Higgs triplets both
acquire non-vanishing
vacuum expectation values $v_{L,R} \neq 0$. This opens up certain decay modes
that are forbidden when $v_L = 0$ (as for a type~I seesaw mechanism) and that
could be used as handles on distinguishing a $v_L = 0$ from a $v_L\neq 0$
scenario. While the phenomenology corresponding to the $v_L\neq 0$ case has been
widely studied in a pure type~II seesaw context~\cite{Han:2007bk,Perez:2008ha,%
delAguila:2008cj,Akeroyd:2010ip,Chun:2003ej,Han:2015hba,Han:2015sca,%
Chabab:2014ara}, the one connected to a mixed type~I/II seesaw model still
remains to be comprehensively explored. Several processes become open only by
virtue of the non-zero $v_L$ value, but still remain suppressed as $v_L$ is
bound to be small. This non-vanishing $v_L$ value nevertheless allows us to
weaken the constraints on the charged scalar particles stemming from
flavour data.

Unlike in earlier MLRSM studies, we consider a high integrated luminosity of LHC
collisions at a centre-of-mass energy of 14~TeV. We adopt
a scenario motivated by current experimental constraints (in particular on
the doubly-charged Higgs bosons) and perform a systematic comparison of expected
new physics signals and Standard Model background including the simulation of
the detector effects.
The rest of this paper is organised as follows. In section~\ref{model}, we
discuss the MLRSM theoretical framework and its particle content, before
designing a representative benchmark scenario that could be probed by
multileptonic probes. In section~\ref{discussion}, we study the collider
phenomenology of this setup and quantitatively estimate how it could be
discovered or constrained at the high-luminosity run of the LHC. We
then generalise our findings as a function of the mass scale of the model,
before concluding in section~\ref{conclusions}.

\section{The Minimal Left-Right Symmetric Model}
\label{model}

\subsection{Theoretical Framework}

\begin{table}
\begin{tabular}{c|c|c|c|c}
Field & $ SU(3) $ & $SU (2)_L$ & $SU(2)_R$ & $U(1)_{B-L}$ \\ 
\hline \hline\\[-0.3cm]
  $q_{L} = \begin{pmatrix}u_{L}\\ d_{L}\end{pmatrix}$
     & $\ {\bf 3}$ & $\ {\bf 2}$ & $\ {\bf 1}$ & $\frac13$
     \\[0.4cm]
  $q_{R} = \begin{pmatrix}u_{R}\\d_{R}\end{pmatrix}$
     & $\ {\bf 3}$ & $\ {\bf 1}$ & $\ {\bf 2}$ & $\frac13$
     \\[0.4cm]
  $\ell_{L} = \begin{pmatrix}\nu_{L}\\e_{L}\end{pmatrix}$
     & $\ {\bf 1}$ & $\ {\bf 2}$ & $\ {\bf 1}$ & $-1$
     \\[0.4cm]
  $\ell_{R}=\begin{pmatrix}\nu_{R}\\e_{R}\end{pmatrix}$
     & $\ {\bf 1}$ & $\ {\bf 1}$ & $\ {\bf 2}$ & $-1$
     \\[0.4cm]
\hline\\[-0.3cm]
  $\Phi= 
   \begin{pmatrix}\phi_1^0 & \phi_1^+ \\ \phi_2^- & \phi_2^0 \end{pmatrix}$
     & $\ {\bf 1}$ & $\ {\bf 2}$ & $\ {\bf 2}$ & $0$
     \\[0.4cm]
  $\Delta_L =
   \begin{pmatrix} \Delta_L^+/\sqrt{2} & \Delta_L^{++} \\
     \Delta_L^0& -\Delta_L^+/\sqrt{2} \end{pmatrix}$
     & $\ {\bf 1}$ & $\ {\bf 3}$ & $\ {\bf 1}$ & $2$
     \\[0.4cm]
  $\Delta_R =
   \begin{pmatrix} \Delta_R^+/\sqrt{2} & \Delta_R^{++} \\
     \Delta_R^0& -\Delta_R^+/\sqrt{2} \end{pmatrix}$
     & $\ {\bf 1}$ & $\ {\bf 1}$ & $\ {\bf 3}$ & $2$
     \\[0.4cm]
\end{tabular}
\caption{\it MLRSM field content, presented together with the representations
   under $SU(3)_c \times SU(2)_L \times SU(2)_R  \times U(1)_{B-L}$.}
\label{table1}
\end{table}

Left-right symmetric models~\cite{Pati:1974yy,Mohapatra:1974hk,Mohapatra:1974gc,
Senjanovic:1975rk,Mohapatra:1977mj,Senjanovic:1978ev,Mohapatra:1980qe,%
Lim:1981kv,Gunion:1989in,Deshpande:1990ip} are Standard
Model extensions where the gauge symmetry group is enlarged to $SU(3)_c \times
SU(2)_L \times SU(2)_R \times U(1)_{B-L}$. In minimal left-right-symmetric
incarnations, the theory is additionally invariant under discrete left-right
symmetry (or $D$-parity) transformations that relate the $SU(2)_L$ and $SU(2)_R$
sectors. Focusing on the minimal model field content shown in
Table~\ref{table1}, the right-handed Standard Model fermionic degrees of freedom
are grouped into $SU(2)_R$ doublets, which renders the presence of right-handed
neutrinos natural. Compared to the Standard Model case, the Higgs sector is
significantly enriched. The Standard Model $SU(2)_L$ Higgs doublet is
promoted to a Higgs $SU(2)_L\times SU(2)_R$ bidoublet $\Phi$ allowing to write
gauge-invariant Yukawa interactions yielding Dirac mass terms for all
fermions, and the breaking of the gauge symmetry down to the
electroweak symmetry further requires the presence of an $SU(2)_R$
triplet $\Delta_R$.  In order to maintain the theory $D$-parity symmetric, we
include its $SU(2)_L$ counterpart $\Delta_L$. All field transformations under a
$D$-parity symmetry operation hence read~\cite{Maiezza:2016ybz}
\be
  \left\{\begin{array}{l}
    q_L \leftrightarrow q_R\\
    \ell_L \leftrightarrow \ell_R\\
    \Delta_L \leftrightarrow \Delta_R\\
    \Phi \leftrightarrow \Phi^\dagger 
  \end{array}\right. \qquad\text{or}\qquad
  \left\{\begin{array}{l}
    q_L \leftrightarrow q_R^c\\
    \ell_L \leftrightarrow \ell_R^c \\
    \Delta_L \leftrightarrow \Delta_R^\ast \\
    \Phi \leftrightarrow \Phi^t
  \end{array}\right. \ ,
\ee
depending whether $D$-parity symmetry is seen as a generalized parity or
charge conjugation.

The Lagrangian of the model is written as~\cite{Alloul:2013fw,Roitgrund:2014zka}
\begin{equation}
  \mathcal{L}_{\text{MLRSM}} = \mathcal{L}_{\text{kinetic}} + 
     \mathcal{L}_{\text{Yukawa}} - V_{\text{scalar}} ,
\label{lagrangian}
\end{equation}
where $\mathcal{L}_{\text{kinetic}}$ contains standard kinetic and gauge
interaction terms for all fields. The Yukawa interactions read
\be\bsp
  \mathcal{L}_{\text{Yukawa}} =&\ -\Big[
     y_{ij}        \bar{\ell}_{iL} \Phi         \ell_{jR}
   + y^\prime_{ij} \bar{\ell}_{iL} \tilde{\Phi} \ell_{jR}
   + Y_{ij}        \bar{q}_{iL} \Phi         q_{jR}
   + Y^\prime_{ij} \bar{q}_{iL} \tilde{\Phi} q_{jR}
   + \frac12 f_{ij}\ \Big(\ell_{iR}^c \tilde \Delta_R \ell_{jR}+
      (R \leftrightarrow L)\Big)+\text{h.c.} \Big]\ ,
\esp\label{treeY}\ee
where $\tilde{\Phi} = \sigma_2\Phi^*\sigma_2$ and
$\tilde\Delta_{L,R} = i \sigma_2 \Delta_{L,R}$. As a consequence of the built-in
$D$-symmetry, both $SU(2)_L$ and $SU(2)_R$ neutrino couplings $f_L$ and $f_R$
are equal to a unique value $f$. The scalar potential $V_{\text{scalar}}$ is
given by
\be\bsp
 V_{\text{scalar}} & = -\mu_1^2 \Tr{\Phi^\dagger\Phi}
   - \mu_2^2\Tr{\Phi^\dagger\tilde{\Phi} + \tilde{\Phi}^\dagger\Phi}
   - \mu_3^2\Tr{\Delta_L^\dagger\Delta_L + \Delta_R^\dagger\Delta_R}
   + \lambda_1\Big(\Tr{\Phi^\dagger\Phi}\Big)^2
 \\ &\
   + \lambda_2\Big\{ \Big(\Tr{\Phi^\dagger\tilde{\Phi}}\Big)^2
       + \Big(\Tr{\tilde{\Phi}^\dagger\Phi}\Big)^2\Big\}
   + \lambda_3 \Tr{\Phi^\dagger\tilde{\Phi}}\Tr{\tilde{\Phi}^\dagger\Phi}
   + \lambda_4\Tr{\Phi^\dagger\Phi}\Tr{\Phi^\dagger\tilde{\Phi}
       + \tilde{\Phi}^\dagger\Phi} \\
  &\
   + \rho_1\Big\{\Big(\Tr{\Delta_L^\dagger\Delta_L}\Big)^2
      + \Big(\Tr{\Delta_R^\dagger\Delta_R}\Big)^2\Big\}
   + \rho_2\Big\{\Tr{\Delta_L\Delta_L}\Tr{\Delta_L^\dagger\Delta_L^\dagger}
      + \Tr{\Delta_R\Delta_R}\Tr{\Delta_R^\dagger\Delta_R^\dagger}\Big\}
  \\&\
   + \rho_3\Tr{\Delta_L^\dagger\Delta_L}\Tr{\Delta_R^\dagger\Delta_R}
   + \rho_4\Big\{\Tr{\Delta_L\Delta_L}\Tr{\Delta_R^\dagger\Delta_R^\dagger}
      + \Tr{\Delta_L^\dagger\Delta_L^\dagger}\Tr{\Delta_R\Delta_R}
  \\&\
   + \alpha_1\Tr{\Phi^\dagger\Phi}
       \Tr{\Delta_L^\dagger\Delta_L + \Delta_R^\dagger\Delta_R}
   + \Big\{
     \alpha_2 \Big(\Tr{\Phi^\dagger\tilde{\Phi}}\Tr{\Delta_L^\dagger\Delta_L}
         + \Tr{\tilde{\Phi}^\dagger\Phi}\Tr{\Delta_R^\dagger\Delta_R}\Big)
      + \text{h.c.}\Big\}   \\&\
   + \alpha_3 \Tr{\Phi\Phi^\dagger\Delta_L\Delta_L^\dagger
      + \Phi^\dagger\Phi\Delta_R\Delta_R^\dagger}
   + \beta_1 \Tr{\Phi^\dagger\Delta_L^\dagger\Phi\Delta_R
      + \Delta_R^\dagger\Phi^\dagger\Delta_L\Phi}
   + \beta_2\Tr{\Phi^\dagger\Delta_L^\dagger\tilde{\Phi}\Delta_R
       + \Delta_R^\dagger\tilde{\Phi}^\dagger\Delta_L\Phi}
   \\ &\
  + \beta_3\Tr{\tilde{\Phi}^\dagger\Delta_L^\dagger\Phi\Delta_R
        + \Delta_R^\dagger\Phi^\dagger\Delta_L\tilde{\Phi}} \ ,
\esp\label{appeneq2}\ee
where we have introduced scalar mass parameters $\mu_i$ and quartic scalar
interaction strengths $\lambda_i$, $\rho_i$, $\alpha_i$ and $\beta_i$.

The symmetry-breaking pattern is split into two steps,
\be
  SU(2)_L \times SU(2)_R \times U(1)_{B-L}
    \ \ \xrightarrow{\langle \Delta_R \rangle} \ \
  SU(2)_L\times U(1)_Y
    \ \ \xrightarrow{\langle \Phi \rangle} \ \
  U(1)_{\rm e.m.} \ .
\label{eq:breaking}\ee
At high energy, the $SU(2)_L\times SU(2)_R
\times U(1)_{B-L}$ symmetry group is first spontaneously broken down to the
electroweak symmetry group, and at a lower energy scale, the electroweak
symmetry is further broken down to electromagnetism. The first breaking step
results from the non-vanishing vacuum expectation value acquired by the
neutral component of the $\Delta_R$ field at the minimum of the scalar
potential, whilst electroweak symmetry breaking is induced by the vacuum
expectation values of the neutral components of the Higgs bidoublet. Introducing
the notations
\be
  \langle \phi^0_{1,2} \rangle = \frac{k_{1,2}}{\sqrt{2}}
  \qquad\text{and}\qquad
  \langle \Delta^0_{L, R} \rangle = \frac{v_{L,R}}{\sqrt{2}}\ ,
\ee
the Standard Model vacuum expectation value is given by $v_{\rm SM} =
\sqrt{k_1^2+k_2^2} \approx 246$~GeV. Without any loss of generality, we make
use of a rotation in the $SU(2)_L\times SU(2)_R$ space so that only one of the
neutral components of the Higgs bidoublet acquires a large vacuum expectation
value, $k_1\approx v_{\rm SM}$
and $k_2\approx0$. In addition, electroweak precision tests constrain $v_L$ to be
smaller than 2~GeV~\cite{Agashe:2014kda}, and the breaking pattern of
Eq.~\eqref{eq:breaking} enforces $v_R$ to be much greater than $k_1$. $D$-parity
invariance moreover imposes the $g_L$ and $g_R$ gauge couplings to be equal to a
common value $g$.

Under those assumption, we
neglect all contributions to the gauge boson masses that are proportional to
$v_L$, so that these masses approximatively read
\be\bsp
  M^2_W= \frac{g^2}{4} k^2_1\ ,
  \quad\quad
  M^2_{W_R} = \frac{g^2}{2}v^2_R\ ,
  \quad\quad
  M^2_Z = \frac{g^2 k^2_1}{4\cos^2{\theta_W}}
   \Big(1-\frac{\cos^2{2\theta_W}}{2\cos^4{\theta_W}}\frac{k^2_1}{v^2_R}\Big)\ ,
  \quad\quad
  M^2_{Z_R} = \frac{g^2 v^2_R \cos^2{\theta_W}}{\cos{2\theta_W}} \ ,
\esp\label{Mgauge}\ee
with $\theta_W$ indicating the weak mixing angle.

After symmetry breaking, the Higgs sector is left with four neutral scalar
fields $H^0_0$, $H^0_1$, $H^0_2$ and $H^0_3$, two neutral pseudoscalar fields
$A^0_1$ and $A^0_2$ (as well as two neutral Goldstone bosons eaten by the $Z$
and $Z_R$ bosons), two singly-charged Higgs bosons $H^{\pm}_1$ and $H^{\pm}_2$
(as well as four charged Goldstone bosons eaten by the $W^\pm$ and $W_R^\pm$
bosons) and two doubly-charged scalar bosons $H^{\pm \pm}_L$ and
$H^{\pm \pm}_R$. Within the approximations above-mentioned, the scalar masses
are given by
\be\bsp
  M^2_{H^0_0} = 2 \lambda_1 k^2_1\ ,
  \qquad
  M^2_{H^0_1} = \frac{1}{2}\alpha_3 v^2_R\ ,
  \qquad
  M^2_{H^0_2} = 2\rho_1 v^2_R\ ,
  \qquad
  M^2_{H^0_3} = \frac{1}{2}(\rho_3-2\rho_1)v^2_R \ ,\\
  M^2_{A^0_1} = \frac{1}{2}\alpha_3 v^2_R - 2(2\lambda_2-\lambda_3)k^2_1\ ,
  \qquad
  M^2_{A^0_2} = \frac{1}{2}v^2_R (\rho_3-2\rho_1)\ ,\\
  M^2_{H^{\pm}_1} = \frac12 (\rho_3-2\rho_1)v^2_R +\frac14 \alpha_3 k^2_1\ ,
  \qquad
  M^2_{H^{\pm}_2} = \frac{1}{2}\alpha_3 v^2_R + \frac{1}{4} \alpha_3 k^2_1\ ,\\
  M^2_{H^{\pm \pm}_L} = \frac12 (\rho_3-2\rho_1)v^2_R+\frac12\alpha_3 k^2_1\ ,
  \qquad
  M^2_{H^{\pm \pm}_R} = 2\rho_2 v^2_R +\frac{1}{2} \alpha_3 k^2_1\ ,
\esp\label{Mscalars}\ee
the left-right triplet mixing induced by the $\beta_i$ potential terms being
neglected as suppressed by the $k^2_1/v^2_R$ ratio, and the different states
being not necessarily mass-ordered.

Turning to the neutrino sector, the $f_{ij}$ Yukawa couplings of
Eq.~\eqref{treeY} give rise to neutrino Majorana masses after symmetry breaking,
as the neutral component of the $\Delta_L$ Higgs field acquires a non-zero
vacuum expectation value~\cite{Deshpande:1990ip},
\be
  v_{L}= \frac{\beta_2 k^2_1}{(2\rho_1-\rho_3) v_R}\ ,
\ee
which satisfies $|v_{L}|\ll v_{\rm SM} \ll |v_{R}|$. The $6\times6$ neutrino
mass matrix is then given, in the $(\nu_L, \nu_R)$ gauge eigenbasis, by
\be
  M = \bpm \sqrt{2} fv_L & M_D \\ M^T_D & M_R \epm \ .
\ee
The light and heavy neutrino sectors are decoupled in the absence of any
left-right mixing potentially induced by a non-vanishing Dirac mass matrix
$M_D$, and the light and heavy masses are in this case respectively governed by
the $f$ and $M_R$ parameters. In contrast, non-zero Dirac masses give rise to
neutrino mixings parameterised by a mixing matrix $R^\nu$,
\be
  R^\nu = \bpm U & S \\ T & V \epm =
   \bpm 1-\frac12 R R^{\dagger} & R \\ -R^{\dagger} & 1-\frac12R^\dagger R\epm
   \bpm U_L & 0 \\ 0 & U_R \epm \ .
\label{mixingmatrix}\ee
In this expression, $R=M_{D} M^{-1}_{R}$, whilst $U_L$ and $U_R$ are
respectively the two matrices diagonalising the light and heavy neutrino mass
matrices $M_\nu$ and $M_R$,
\be
  (M_{\nu})_{ij} = \sqrt{2}f_{ij} v_L - (M_D)_{ik} (M_R)^{-1}_{kl} (M^T_D)_{lj}
  \qquad\text{with}\qquad
  (M_D)_{ij} = \frac{1}{\sqrt{2}}y_{ij}k_1
   \quad\text{and}\quad
  (M_R)_{ij} = \sqrt{2}f_{ij} v_R\ .
\ee
This shows that light neutrino masses arise from a combination of type-I and
type-II seesaw contributions and are derived from the diagonalisation of the
upper-left block of the mass matrix $M$ by a $U_L$ rotation, $U_L$ being the
usual Pontecorvo-Maki-Nakagawa-Sakata (PMNS) matrix.

\subsection{Constraints and MLRSM Benchmark Scenarios with Light Doubly-Charged
  Higgs Bosons}
\label{constraints}
In order to design simplified phenomenologically viable benchmark scenarios for
the collider studies performed in the next section, we account for various
constraints arising from current data.

The scalar sector of the theory must include a neutral scalar boson that is
consistent with the observation of a Standard-Model-like Higgs boson with a mass
of about 125~GeV. We enforce the $H^0_0$ boson to be such a boson, its
mass being set to
\be
  M_{H_0^0} = 125~{\rm GeV.}
  \label{eq:SMhiggs}
\ee
The extra neutral scalar bosons, in particular those with a large bidoublet
component ({\it i.e.}, $H^0_1$ and $A^0_1$), generally mediate tree-level
flavor-changing
neutral interactions. Consequently, their mass is constrained to be above about
10~TeV~\cite{Ecker:1983uh,Mohapatra:1983ae,Pospelov:1996fq,Zhang:2007da,%
Maiezza:2010ic} from experimental kaon mixing data~\cite{Beall:1981ze}. On the
other hand, the scalar potential perturbativity and unitarity further
push these Higgs bosons to be heavier than 18~TeV. The minimum mass
configuration is realised for a $W_R$ boson satisfying
$M_{W_R}>8$~TeV~\cite{Maiezza:2016bzp}, a constraint that is by far compatible
with the most stringent LHC bounds regardless of the details of he right-handed
neutrino sector~\cite{Aaboud:2017efa,Aaboud:2017yvp,Sirunyan:2016iap,%
Khachatryan:2016jww,Sirunyan:2018mpc}. Consequently, we impose
\be
  M_{H_1^0} = M_{A_1^0}=20~{\rm TeV,}\qquad
  M_{W_R} = 10~{\rm TeV}\ \ \Leftrightarrow\ \
  v_{R} = 21639.39~{\rm GeV}.
\label{eq:MWR}\ee
Saturating the present limits on the $SU(2)_L$ triplet vacuum expectation
value arising from the $\rho$-paramater~\cite{Agashe:2014kda},
\be
  v_L = 2~{\rm GeV,}
\ee
and recalling that we have chosen
\be
  k_{1} \approx v_{\rm SM} = 246~{\rm GeV}\qquad\text{and}\qquad  k_{2} \approx 0,
\ee
we make use of Eq.~\eqref{Mscalars} together with the setup of Eq.~\eqref{eq:SMhiggs} through
Eq.~\eqref{eq:MWR} to numerically derive the (tree-level) parameters of
the scalar potential,
\be
  \lambda_1 = 0.129\ , \qquad
    \alpha_{3} = 1.708\ , \qquad
    2\lambda_2-\lambda_3 = 0.
\ee
On the basis of the results of the ATLAS searches for same-sign dileptonic new
physics signals~\cite{Aaboud:2017qph}, we moreover impose a lower bound
on the masses of the doubly-charged
scalars $H^{\pm \pm}_L$ and $H^{\pm \pm}_R$. Assuming that the branching
ratios into electronic and muonic final states are both equal to 50\%, the
$SU(2)_L$ and $SU(2)_R$ doubly-charged Higgs-boson masses have to be larger than
785~GeV and 675~GeV respectively. We adopt an optimistic scenario and take their
masses close to the experimental limits,
\be
  M_{H^{++}} \equiv M_{H^{++}_L}= M_{H^{++}_R}= 800~{\rm GeV.}
\label{eq:Hppmasses2}\ee
This leads to
\be
     \rho_{2} = 6.2817 \times 10^{-4}\ , \qquad
      \rho_{3}-2\rho_1 = 0.0025 \ ,
\ee
with the other gauge and scalar boson masses being therefore
\be
  M_{Z_{R}} = 16754~{\rm GeV}, \quad
  M_{H^{0}_{2}} = 6756~{\rm GeV},\quad
  M_{H^{0}_{3}} = M_{A^0_2} = 767~{\rm GeV},\quad
  M_{H_{1}^{+}} =  784~{\rm GeV},\quad
  M_{H_{2}^{+}}= 20 ~{\rm TeV}.
\ee
The parameters related to the neutrino masses and mixings can be constrained by
LHC searches in the same-sign dilepton plus dijet channel, such a signature
being relevant for probing right-handed neutrino production via an $s$-channel
$W_R$ exchange~\cite{Keung:1983uu}. A wide fraction of the parameter space turns
out to be excluded by 8~TeV and 13~TeV LHC data due to the non-observation of
any such signal~\cite{Basso:2015pka,Sirunyan:2018pom}. For the choice of
Eq.~\eqref{eq:MWR}, right-handed neutrinos have to be heavy.
The neutrino sector can also undergo several low energy tests from intensity
frontier experiments looking for lepton-number violation (such as
neutrinoless double-beta decays) or for lepton-flavor violation (like rare
muonic decays into electrons such as $\mu \to e \gamma$ or $\mu\to 3e$).
The associated combined limits induce a hierarchy between the mass of the
$SU(2)_R$ scalar bosons and the mass of the heaviest right-handed neutrino that
must be 2 to 10 times smaller~\cite{Tello:2010am,Bambhaniya:2015ipg} for
$M_{W_R}=3.5$~TeV.
These bounds are however derived under the assumption that either a type I or a
type II seesaw mechanism is implemented.
Considering a model featuring a combination of type I and type II seesaw
mechanisms (as in this work) enables us to evade those
bounds~\cite{Borah:2015ufa,Borah:2016iqd}, the $SU(2)_R$ triplet scalar masses
being even allowed to be smaller than the heaviest right-handed neutrino mass.
Right-handed neutrinos could nevertheless be indirectly constrained by
neutrinoless double-beta decays and cosmology~\cite{Borah:2016lrl,Frank:2017tsm,
Araz:2017qcs}.

With the above assignments, the only parameters left to be considered are the
mass parameters of the heavy and light neutrinos, $M_{N_i}$, with $i=1$, 2, 3,
4, 5, $6$.  We shall consider the lighter degrees of freedom ({\it i.e.}, the
left-handed neutrinos corresponding to $i=1$, 2, 3) to have a mass of the order
of $0.1$~eV to agree with cosmological data. For simplicity, we assume a
unified scenario for the right-handed neutrino sector,
\be
  M_{N_4} = M_{N_5} = M_{N_6} \equiv M_{N_R} =12~{\rm TeV} \ ,
\ee
which allows one to evade all the above-mentioned bounds and feature
perturbative Yukawa couplings of ${\cal O}(1)$.

In addition, the heavy-light neutrino mixing is constrained from neutrino
oscillation data. After having fixed all physical masses of the neutrinos
and assuming that $|f v_L| \ll |M_D| \ll |M_R|$, the $M_R$ mass matrix can be
read off these masses, in a first approximation, and the type-II seesaw
contribution to the light neutrino masses can be deduced from the (inputted)
$M_\nu$ matrix~\cite{Nemevsek:2012iq},
\be
  M_{D} = M_{N_R} \left ( \frac{v_L}{v_R}-\frac{M_{\nu}}{M_{N_R}}  \right )^{1/2} \  \ .
\label{eqmLR}\ee
The Dirac mass contribution is of about 100~GeV with an assumed $M_\nu\sim 0.1$
eV for the considered scenario. The cancellation in Eq.~\eqref{eqmLR} between
the Type I and Type II seesaw contributions are hence fine-tuned to the level of
$10^{10}$, such a fine-tuning being stable against quantum
corrections~\cite{Luo:2008rs}.
This further impacts the light-heavy neutrino mixing matrix,
\be
  T =-R^\dagger U_L = -(M^{-1}_R)^\dagger M^\dagger_D U_L \approx 0.018 U_L\ ,
\ee
which also dictates the strength of the heavy neutrino decays into left-handed
leptons and a Standard Model $W$-boson. The rest of the neutrino mixing matrix
stems from the unitarity properties of that matrix, which subsequently
fixes the strength of the heavy neutrino decays into a final state system made
of a right-handed lepton and a possibly off-shell heavy gauge boson.
The heavy-light neutrino mixing is actually rather large, as
$v_L$ is large, which opens the door for unusual heavy
neutrino decays into the left-handed sector.

\subsection{Main Feature of our Benchmark Scenario}
\begin{table}
\renewcommand{\arraystretch}{1.4}
\setlength\tabcolsep{5pt}
\begin{tabular}{l|r}
\multicolumn{2}{c}{Doubly-charged bosons}\\  \hline\hline
    BR$\big(H_R^{++} \to \ell^{+} \ell^{+}\big)$ & 33.3\% \\
  \hline
    BR$\big(H_L^{++} \to \ell^{+} \ell^{+}\big)$ & 32.39\% \\
    BR$\big(H_L^{++} \to W^{+} \ell^+ \nu_{\ell} \big)$ & 1.00\%\\
\multicolumn{2}{c}{~}\\
\multicolumn{2}{c}{Singly-charged bosons}\\  \hline\hline
    BR$\big(H_1^+ \to \ell^{+} \nu_\ell\big)$ &33.26\% \\
    BR$\big(H_1^+ \to W^{+} Z \big)$ & 0.22\% \\
  \hline
    BR$\big(W_R^+ \to q \bar q'\big)$  & 100\% \\
 \end{tabular}
\hspace{0.9cm}
\begin{tabular}{l|r}
\multicolumn{2}{c}{Neutral bosons}\\ \hline\hline
    BR$\big(H_3^0 \to \nu_\ell \nu_\ell\big)$ & 31.8\% \\
    BR$\big(H_3^0 \to Z  \nu_\ell \nu_\ell\big)$ & 0.83\%\\
    BR$\big(H_3^0 \to W \ell \nu_\ell\big)$ & 0.38\%\\
    BR$\big(H_3^0 \to Z  Z\big)$ & 0.52\% \\
    BR$\big(H_3^0 \to W W \big)$ & 0.26\% \\
  \hline
    BR$\big(A_2^0 \to \nu_\ell \nu_\ell\big)$   & 32.1\%\\
    BR$\big(A_2^0 \to Z \nu_\ell \nu_\ell\big)$ & 0.85\%\\
    BR$\big(A_2^0 \to W \ell \nu_\ell\big)$ & 0.39\%\\
 \end{tabular}
\caption{\it Branching ratios (BR) associated with the different decay channels
  of the light scalar and vector bosons within the considered MLRSM realisation.
  We independently denote by $\ell = e, \mu, \tau$ any lepton flavor, and omit
  any channel whose branching ratio is smaller than 0.1\%.}
\label{tab:bp2_decays}
\end{table}

In order to determine the experimental signatures associated with the production
of two scalar bosons or one vector and one scalar boson at the LHC, we
first present, in Table~\ref{tab:bp2_decays}, the decay table related to the
relevant (lighter) fields. The large right-handed neutrino masses have
deep consequences on the couplings of the $SU(2)_L$ doubly-charged and
singly-charged Higgs bosons to leptons, as they are proportional to
$M_{N_R}/v_R$. This impacts the decay pattern of the scalar fields
which will rarely decay into non-leptonic final states. The $H_L^{\pm\pm}$ boson
hence almost decays exclusively into a same-sign dileptonic system and the
$H_1^+$
Higgs boson into a lepton-neutrino pair. Other non negligibly small $H_L^{++}$
decay modes include a virtual $H_1^+$ boson,
and these sum up to a branching ratio of
3\% after considering all three lepton flavours. The $SU(2)_R$ doubly-charged
$H_R^{++}$ boson in contrast only decays into a same-sign lepton pair, the
potential decay modes into an $SU(2)_R$ gauge boson and another scalar being
kinematically closed. Finally,
the lighter neutral $H_3^0$ and $A_2^0$ bosons almost exclusively decay
invisibly, the neutral scalar boson $H_3^0$ nevertheless undergoing rare visible
decays into electroweak gauge bosons, whilst the singly-charged $W_R$ boson
always decays into a dijet system.

\begin{figure}
\centering
  \includegraphics[height=3.0cm]{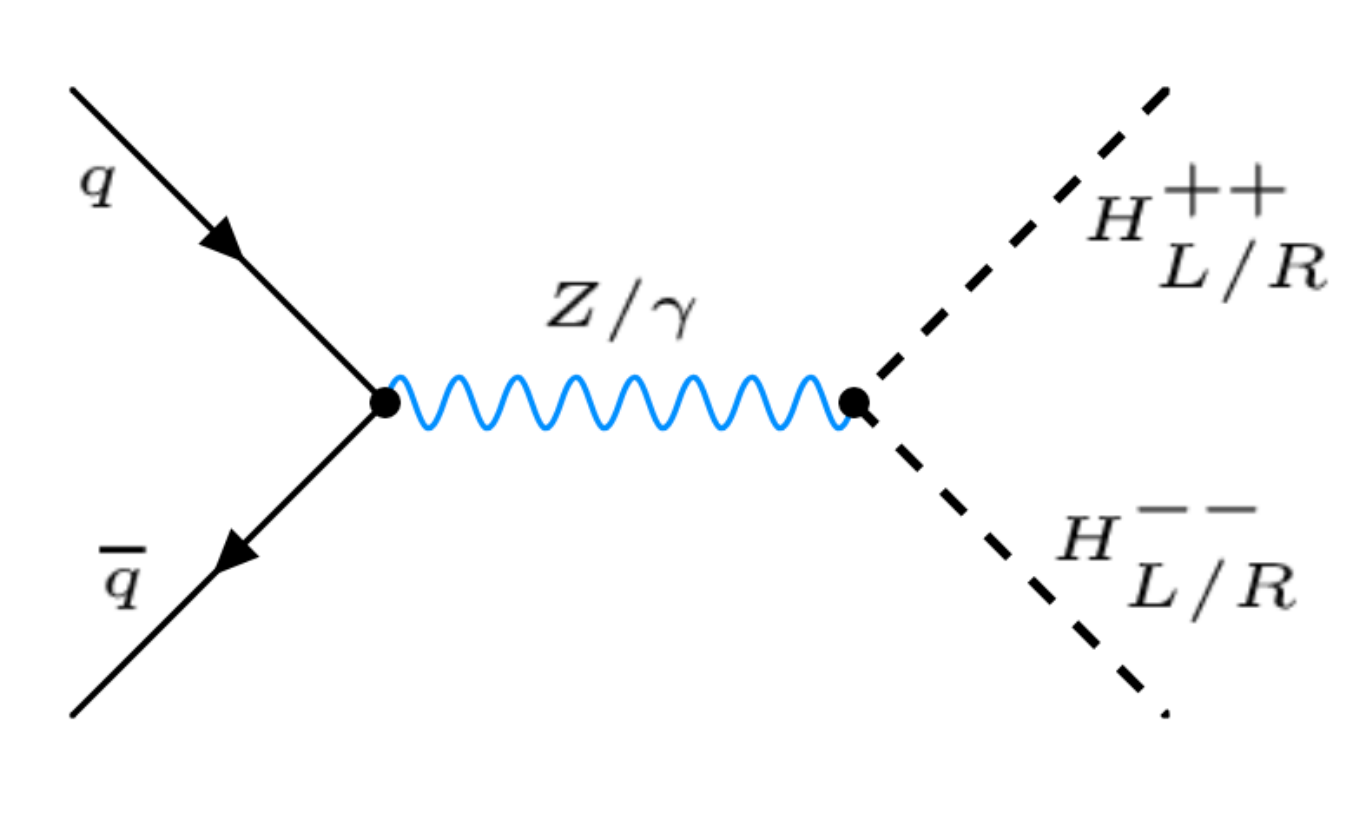}
  \includegraphics[height=3.0cm]{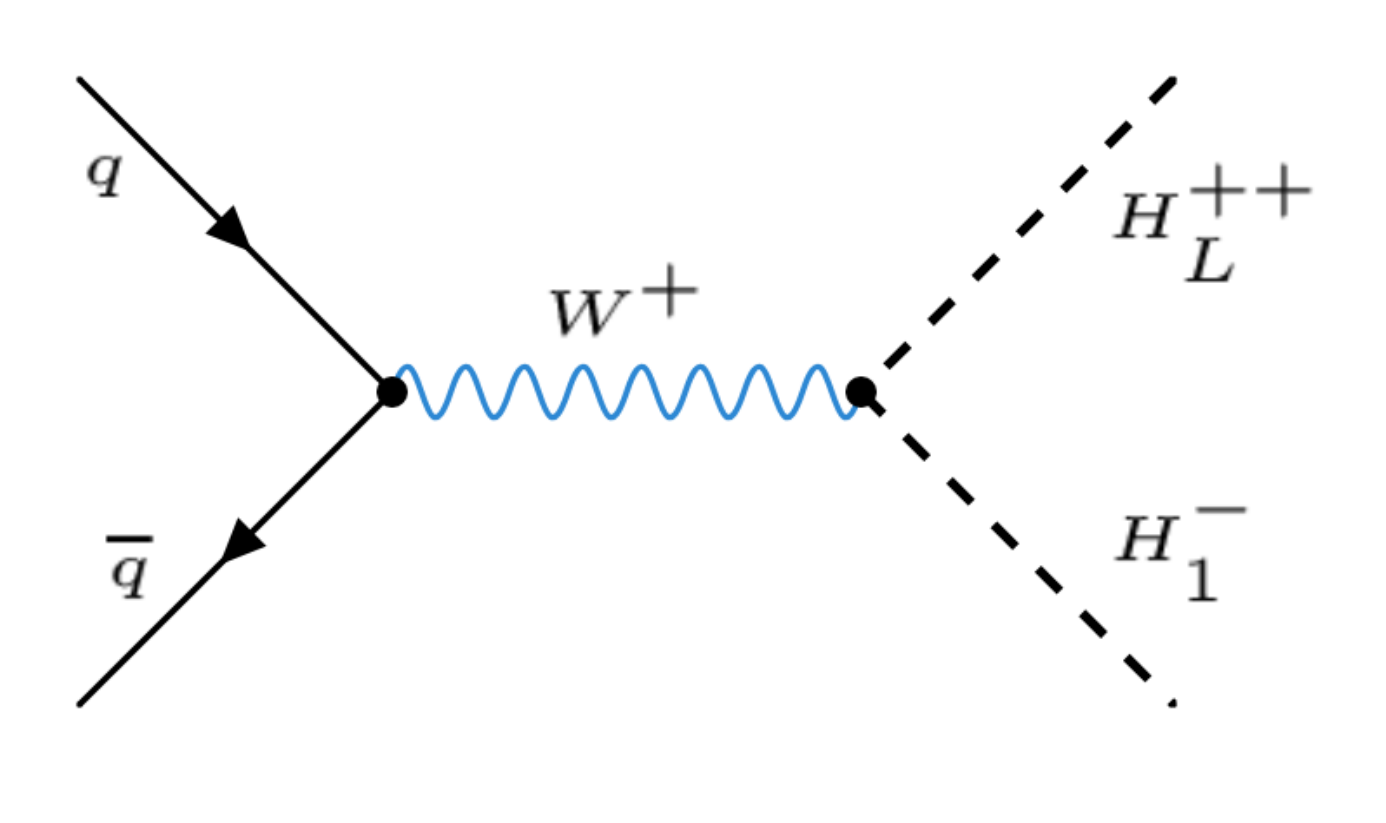}
  \includegraphics[height=3.0cm]{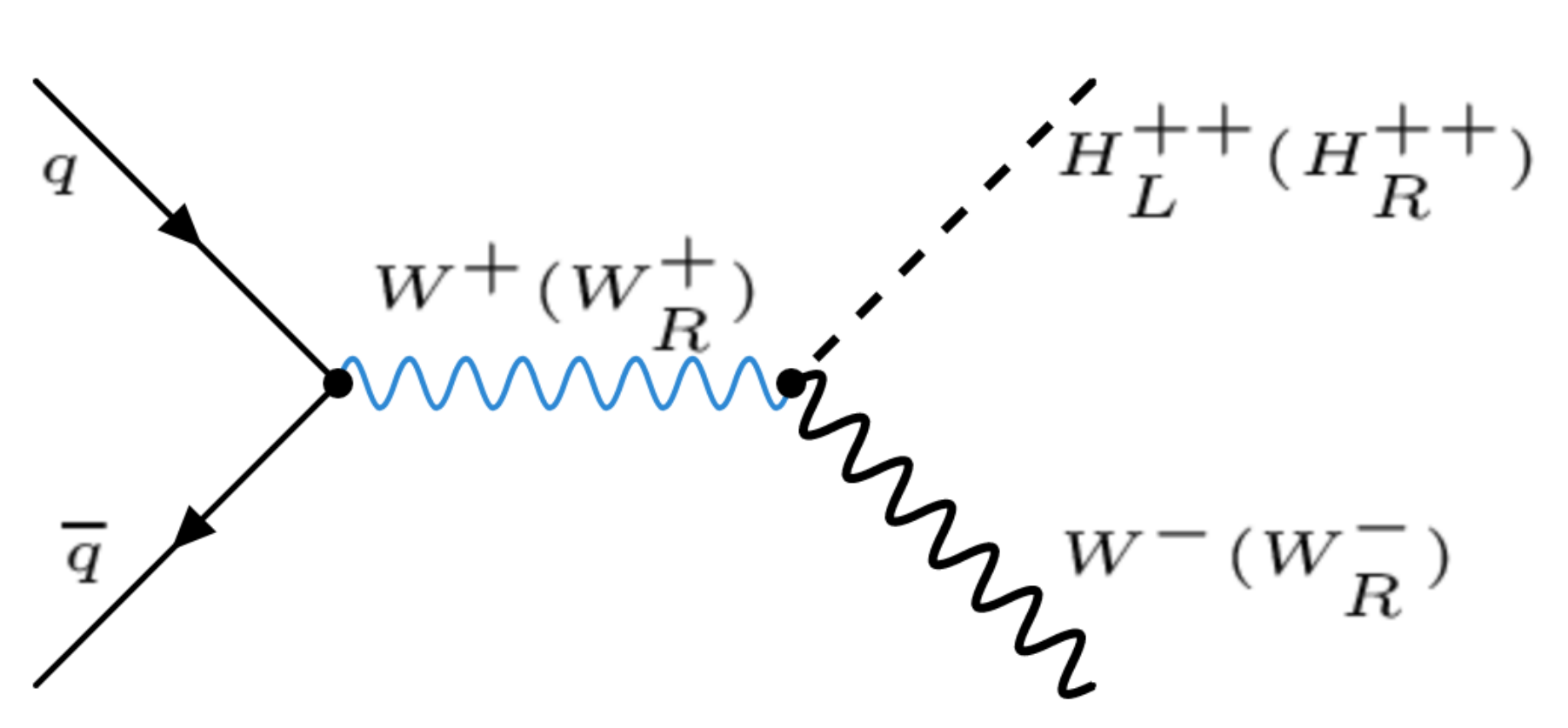}
  \caption{\it Representative Feynman diagram of the various MLRSM processes
    giving rise to multileptonic final states. }
\label{fig:diagrams}
\end{figure}

As the considered new particles significantly decay into leptonic final
states, natural collider probes include final-state systems made of three or
four leptons, as the corresponding Standard Model background is additionally
small. As a consequence, we focus on the production of two
doubly-charged Higgs bosons and on the associated production of a doubly-charged
Higgs boson and a singly-charged gauge or Higgs boson,
\be
  p p \to H^{++}_{L/R} H^{--}_{L/R}\ , \qquad
  p p \to H^{\pm\pm}_{L} H^{\mp}_1\qquad\text{and} \qquad
  p p \to H^{\pm\pm}_{R} W^\mp_R\ ,
\label{eq:process1}\ee
for which representative leading-order
Feynman diagrams are shown in Figure~\ref{fig:diagrams}. Other hard-scattering
processes involving new Higgs and gauge bosons could also possibly lead to
multileptonic final states, but with suppressed and negligible rates. For
instance, $H_L^{\pm\pm}W^\mp$ production suffers from a
strong $v_L$ suppression, and the large mass of the heavier Higgs bosons yields
to a severe phase-space suppression for any process in which they could be
produced.

\begin{table}
\renewcommand{\arraystretch}{1.7}
\setlength\tabcolsep{5pt}
\begin{tabular}{l|r}
\multicolumn{2}{c}{Total rate at $\sqrt{s}=14$~TeV }\\
  \hline\hline
    $pp \to H_L^{++}H_L^{--}$ &  0.197~fb    \\
    $pp \to H_R^{++}H_R^{--}$ &  0.076~fb    \\
  \hline
    $pp \to H_L^{++} H_1^-$ & 0.28~fb \\
    $pp \to H_L^{--} H_1^+$ & 0.10~fb \\
\end{tabular}
\hspace{0.5cm}
\begin{tabular}{l|r}
\multicolumn{2}{c}{Total rate at $\sqrt{s}=14$~TeV }\\
  \hline\hline
    $pp \to H_{L/R}^{++}H_{L/R}^{--} \to
      \ell^+\ell^+\ell^-\ell^-$ & 0.12~fb \\
  \hline
    $pp \to H_L^{\pm\pm}H_{1}^{\mp} \to
      \ell^\pm\ell^\pm\ell^\mp + \met\ $ & 0.17~fb\\
\end{tabular}
\caption{\it Production cross sections associated with the set of processes
  shown in Eq.~\eqref{eq:process1}, for proton-proton collisions at a
  centre-of-mass energy of 14~TeV and in the context of the adopted benchmark
  scenario. The cross sections are obtained by multiplying the results returned
  by MG5\_aMC~\cite{Alwall:2014hca}, when leading-order matrix elements are
  convoluted with the leading-order set of NNPDF~2.3 parton
  densities~\cite{Ball:2012cx}, with an NLO $K$-factor of
  1.25~\cite{Muhlleitner:2003me}. Total
  production rates are presented on the left panel, whereas branching ratios
  into the two final states of interests are included on the right panel. We
  independently denote by $\ell = e, \mu$ any light lepton flavour (and the
  lepton flavours can be different within any given process).}
\label{tab:bp2_xsections}
\end{table}

Leading-order cross sections are given in the top panel of
Table~\ref{tab:bp2_xsections} for LHC proton-proton collisions at a
centre-of-mass energy $\sqrt{s}=14$~TeV.
The production of a pair of doubly-charged Higgs
bosons proceeds via a Drell-Yan-like process and the exchange of an $s$-channel
neutral gauge or Higgs boson. By virtue of the smallness of the Standard Model
Yukawa couplings and the heavy mass of the $Z_R$ and extra Higgs bosons, virtual
$Z$-boson and photon contributions dominate and lead to a cross section of about
0.197~fb and 0.076~fb for the production of a pair of $SU(2)_L$ and $SU(2)_R$
doubly-charged Higgs bosons respectively, including in both case a
next-to-leading order (NLO) $K$-factor of 1.25~\cite{Muhlleitner:2003me}.
Doubly-charged Higgs bosons can also
be produced in association with a singly-charged Higgs or gauge boson. While the
$H_R^{\pm\pm}W_R^\mp$ production cross section is negligible by virtue of the
heavy mass of the $W_R$ boson, a $H_L^{\pm\pm}H_{1}^{\mp}$ system can be
produced via the exchange of a lighter Standard Model $W$-boson, other diagrams
contributing to a smaller extent. The associated cross section is of about
0.38~fb.

Including the relevant branching ratios,
four-lepton final states arise from the production and decay of a pair
of doubly-charged Higgs bosons each decaying into a same-sign dilepton,
the associated cross section being of about 0.12~fb. On the other hand,
trileptonic final states originate from the associated production of a
doubly-charged and a singly-charged Higgs boson, with a similar total rate of
0.17~fb. Other leptonic final states, like those featuring a same-sign dilepton
system, cannot be produced with a sufficiently large rate to be relevant.
Despite the smallness of these cross sections, we demonstrate in
Section~\ref{discussion} that the associated MLRSM signals can potentially be
observed (or excluded) at the high-luminosity run of the LHC.

\section{LHC phenomenology}
\label{discussion}
For our analysis, we have used the {\sc FeynRules} package~\cite{%
Alloul:2013bka} and the existing implementation of the MLRSM model~\cite{%
Roitgrund:2014zka} to generate a UFO model~\cite{Degrande:2011ua} that can be
used within the MG5\_aMC platform~\cite{Alwall:2014hca}. We have generated
hard-scattering events both for the signal processes of
Eq.~\eqref{eq:process1} and for the Standard Model
background, the tree-level matrix elements being convoluted with
the leading-order set of NNPDF~2.3 parton distributions~\cite{Ball:2012cx}. The
simulation of the QCD environment (parton showering and hadronisation) has been
performed with {\sc Pythia 6}~\cite{Sjostrand:2006za}, and we
have included the response of a CMS-like detector with {\sc Delphes
3}~\cite{deFavereau:2013fsa} that internally relies on
{\sc FastJet}~\cite{Cacciari:2011ma} for the reconstruction of the physics
objects, using the anti-$k_T$ algorithm with a radius parameter $R=
0.4$~\cite{Cacciari:2008gp}.

We require, at the matrix-element level, that all leptons and jets have a
transverse momentum $p_T^{\rm gen}$ and pseudorapidity $\eta^{\rm gen}$
satisfying
\be
  p_T^{\rm gen}(j) > 10~{\rm GeV}, \qquad
  p_T^{\rm gen}(\ell) > 10~{\rm GeV}, \qquad
  |\eta^{\rm gen}(j)| < 2.5 \qquad\text{and}\qquad
  |\eta^{\rm gen}(\ell)| < 2.5 \ ,
\ee
and are separated in the transverse plane by an angular distance of at least
0.4,
\be
  \Delta R^{\rm gen}(j,j) > 0.4\ , \qquad
  \Delta R^{\rm gen}(j,\ell) > 0.4 \qquad\text{and}\qquad
  \Delta R^{\rm gen}(\ell,\ell) > 0.4\ .
\ee
We analyse the reconstructed events with {\sc MadAnalysis~5}~\cite{Conte:2012fm}
and impose a basic event preselection where the reconstructed leptons and jets
are required to be central and to have a transverse momentum larger than 20~GeV,
\be
  p_T(j) > 20~{\rm GeV},\qquad
  p_T(\ell) > 20~{\rm GeV}, \qquad
  |\eta(j)| < 2.5 \qquad\text{and}\qquad
  |\eta(\ell)| < 2.5  \ .
\ee
We furthermore ignore any lepton lying within a cone of radius $R=0.4$ centred
on a jet.

\subsection{Four-Lepton Probes}
\label{sec:4l}
After simulating all the potential contributions to the Standard Model
background (except for fake and charge-misidentificaton contributions), our
preselection implies that the main background to a four-lepton signal are
events issued from the production of a pair of (possibly off-shell) $Z$-bosons
where
both weak bosons decay leptonically. Subdominant contributions are expected to
originate from the production of a $WWZ$ system and the associated production of
a top quark-antiquark pair with a $Z$-boson. These last components of the
background could potentially be rejected by (at least loosely) vetoing the
presence of missing energy and $b$-tagged jets in the final state. The signal
fiducial cross section, normalised at the NLO accuracy and including the basic
preselection is of 0.11~fb, for a
corresponding background cross section of 18.9~fb, 1.3~fb and 0.13~fb for
the $Z$-boson pair, $t\bar t Z$ and tribosonic components, respectively. These
last numbers include a QCD next-to-next-to-leading-order (NNLO) $K$-factor of
1.72 for the diboson case~\cite{Cascioli:2014yka} and NLO ones of 1.38 and 1.04
for the two other
processes~\cite{Kardos:2011na,Alwall:2014hca} respectively.

\begin{figure}
  \centering
  \includegraphics[width=0.49\columnwidth]{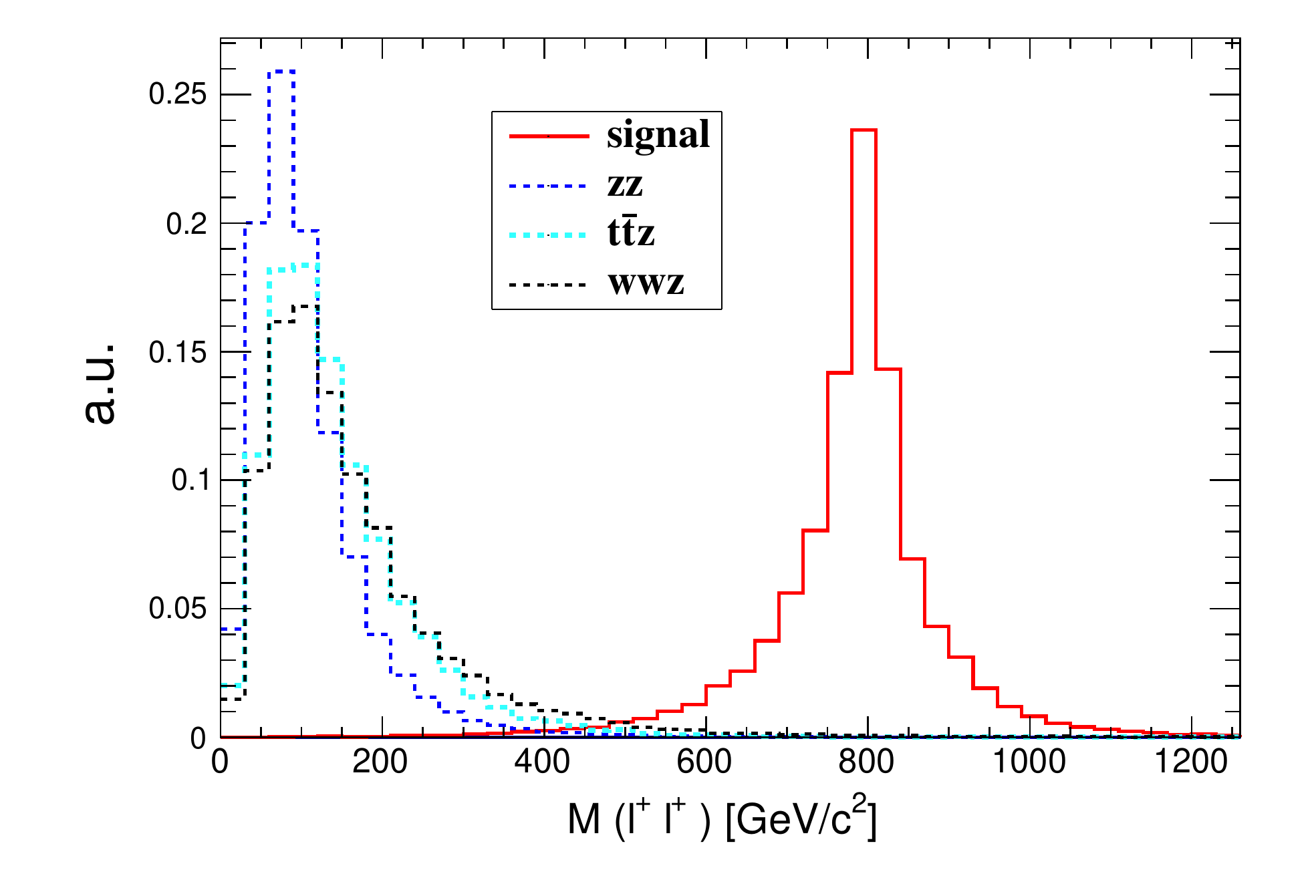}
  \caption{\it Normalized invariant mass spectrum of the system made of the two
    positively-charged leptons, after selecting events containing two pairs of
    same-sign leptons and vetoing the presence of $b$-tagged jets for the signal
    and the background.}
\label{fig_m2l}
\end{figure}

To optimise the signal significance, we select events featuring exactly two
pairs of opposite-charge leptons and veto those exhibiting any $b$-tagged jet.
Signal leptons being originating from the decay of heavy Higgs bosons, we
further impose that the $p_T$ of the three leading leptons $p_T(\ell_1)$,
$p_T(\ell_2)$ and $p_T(\ell_3)$ satisfy
\be
  p_T(\ell_1) > 200~{\rm GeV},\qquad
  p_T(\ell_2) > 150~{\rm GeV}\qquad\text{and}\qquad
  p_T(\ell_3) > 60~{\rm GeV}.
\ee
We then use, as an extra handle on the new physics signal, the
invariant masses of the systems formed by the two pairs of same-sign leptons
$M({\ell^+\ell^+})$ and $M({\ell^-\ell^-})$. As
in any resonance search, the shape of the signal spectrum is expected to show
peaks corresponding to the physical masses of the parent particles, \ie,
the doubly-charged Higgs bosons in our case, as illustrated in
Figure~\ref{fig_m2l}. By imposing that these invariant masses fulfil
\be
  M(\ell^+\ell^+) > 300~{\rm GeV}\qquad\text{and}\qquad
  M(\ell^-\ell^-) > 300~{\rm GeV},
\ee
we are able to make the selection almost free from any background
contamination, with about 15 background events being expected for a luminosity
of 1~ab$^{-1}$.

\begin{figure}
  \centering
  \includegraphics[width=0.46\columnwidth]{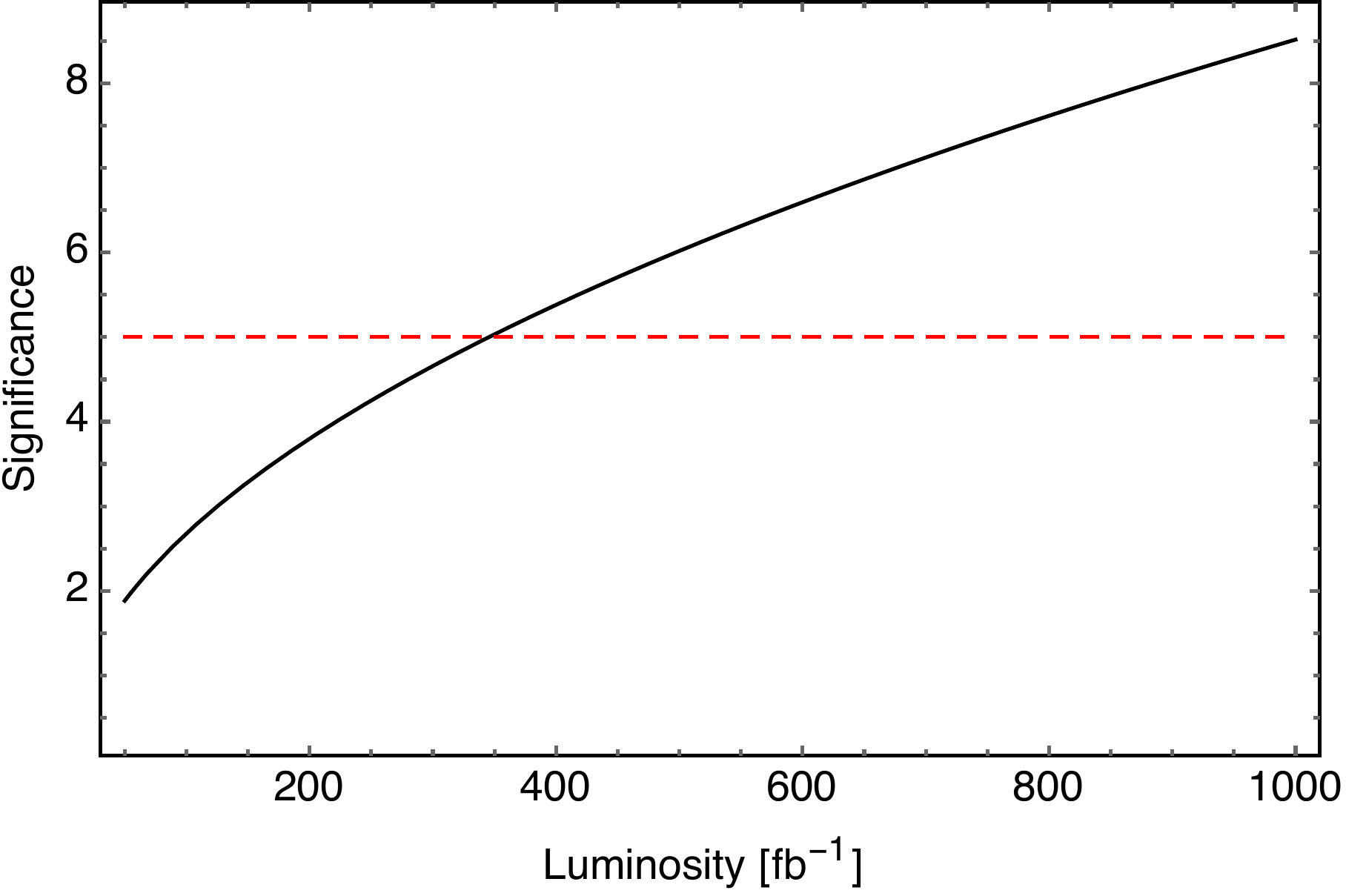}
  \hspace*{.7cm}\includegraphics[width=0.48\columnwidth]{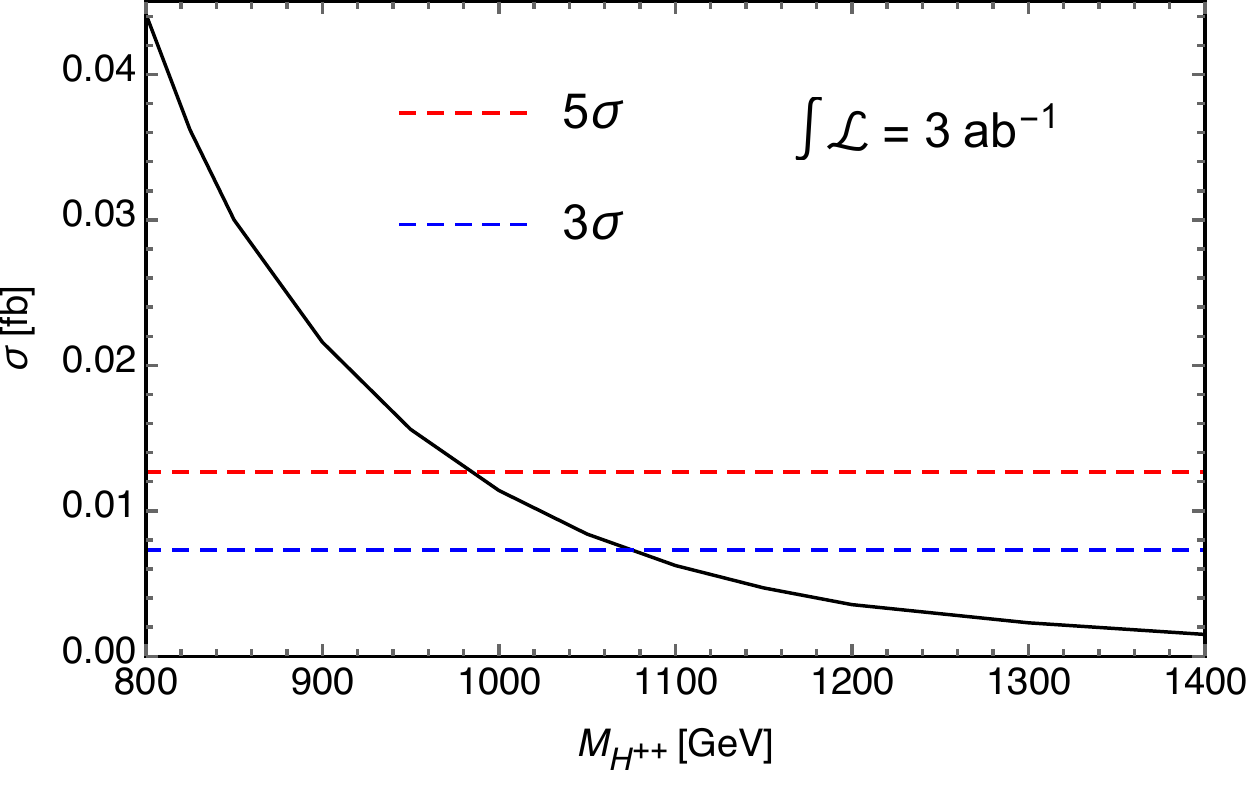}
  \caption{Left: {\it Dependence of the LHC sensitivity to the
    four-leptonic MLRSM signal on the integrated luminosity.} Right: {\it
    Dependence of the signal fiducial cross section after all selections on the
    double-charged Higgs boson mass $M_{H^{++}}$. The corresponding $3\sigma$
    (blue) and $5\sigma$ (red) reference lines are indicated, assuming an
    integrated luminosity of 3~ab$^{-1}$.}}
\label{fig_4l-obs}
\end{figure}

In the left panel of Figure~\ref{fig_4l-obs}, we present the sensitivity of the
LHC to the MLRSM four-leptonic signal for different luminosity goals, the
sensitivity $s$ (expressed in $\sigma$) being defined by ~\cite{Cowan:2010js}
\begin{equation}
 s = \sqrt{2}~\times\left((S+B)~\ln\left[\frac{(S+B)~(B+x^2)}{B^2+(S+B)~x^2}\right] - \frac{B^2}{x^2}\ln\left[1+\frac{x^2~S}{B~(B+x^2)}\right]\right)^{\frac{1}{2}} \ ,
\label{eq:sign}\end{equation}
where $S$ and $B$ respectively indicate the number of surviving signal and
background events, and $x$ represents the systematic uncertainties on the
background. Assuming $x=0.1 B$, a discovery could occur for about 400~fb$^{-1}$,
whilst a signal significance of $s=8.5 \sigma$ could be expected for
1~ab$^{-1}$. Conversely, assuming that exclusion statements could be achieved
for a sensitivity of about $2\sigma$, the considered benchmark scenario could
be excluded at the very beginning of the LHC Run~3. In order
to assess the stability of our predictions with respect to the systematics, we
vary the $x$ parameter to up to 20\% of the background, and investigate the
induced modifications on the predictions. The latter are found robust and almost
agnostic of such a change.
In right panel of Figure~\ref{fig_4l-obs}, we generalise our conclusions to
heavier scenarios and present the dependence of the fiducial cross
section associated with the production of the four-lepton signal on the mass of
the doubly-charged Higgs boson. We compare it to the cross sections yielding
potential $3\sigma$ (blue) and $5\sigma$ (red) observations for an integrated
luminosity of 3~ab$^{-1}$. Whilst the discovery reach can in principle be
pushed above 1.1~TeV regime, it is clear that hints for new physics could be
observed for much heavier scales. Four-leptonic probes consist thus of key MLRSM
signals, both by virtue of the associated background-free environment and by the
moderate value of the signal cross sections.

\subsection{Trileptonic Probes}

With a larger signal cross section, trileptonic probes are expected to provide
good handles on any potential new physics signal. In our case, trileptonic
signal events originate from the associated production of a doubly-charged and
a singly-charged scalar. The signal production rate after preselection is
of about 0.14~fb. Such a signal rate must compete with diboson and triboson
backgrounds as well as with top backgrounds through the associated production of
a top-antitop pair with a weak boson, any other background contribution having
been found negligible after requesting the presence of three leptons. The
corresponding cross sections are of 214.5~fb,
32.5~fb, 1~fb, 2.77~fb and 1.71~fb for $WZ$, $ZZ$, $WWW$, $ttZ$ and $ttW$
production, after including NNLO QCD $K$-factors in the diboson cases
($K_{WZ}=2.01$ and $K_{ZZ}=1.72$~\cite{Grazzini:2016swo,Cascioli:2014yka}), an
NLO QCD and electroweak $K$-factor in the triboson case
(we have conservatively chosen $K_{WWW}=2.27$~\cite{Yong-Bai:2016sal}, which
differs from the much smaller $K$-factor of Ref.~\cite{Campanario:2008yg}),
an NLO QCD $K$-factor in the $ttZ$
case ($K_{ttZ} = 1.38$~\cite{Kardos:2011na}) and an NLO $K$-factor including
the resummation of the threshold logarithms in the $ttW$ case ($K_{ttW} =
1.07$~\cite{Li:2014ula}).

Our selection requires events to contain two leptons carrying the same electric
charge and a third lepton with an opposite charge. We moreover veto the presence
of any reconstructed $b$-tagged jet to control the top-quark-induced background.
As in the previous section, we impose a stringent selection of the lepton
properties, and constrain the transverse-momentum of the two leading leptons to
satisfy
\be
  p_T(\ell_1) > 250~{\rm GeV} \qquad\text{and}\qquad
  p_T(\ell_2) >  80~{\rm GeV}.
\ee
The signal selection efficiency is practically of 100\%, while the
dominant diboson background is reduced by a factor of 50. We then reconstruct
the invariant mass of the same-sign dilepton system and use it as an extra
handle on the signal, requiring
\be
  M_{\ell^\pm\ell^\pm} > 300~{\rm GeV},
\ee
which reduces
the triboson and remaining top-induced background to a barely visible level.
Although a veto on events featuring an opposite-sign dilepton system compatible
with a $Z$-boson could help in reducing the remaining background, we instead
require the selected events to contain a large amount of missing transverse
energy,
\be
  \met > 150~{\rm GeV}.
\ee

\begin{figure}
  \centering
  \includegraphics[width=0.47\columnwidth]{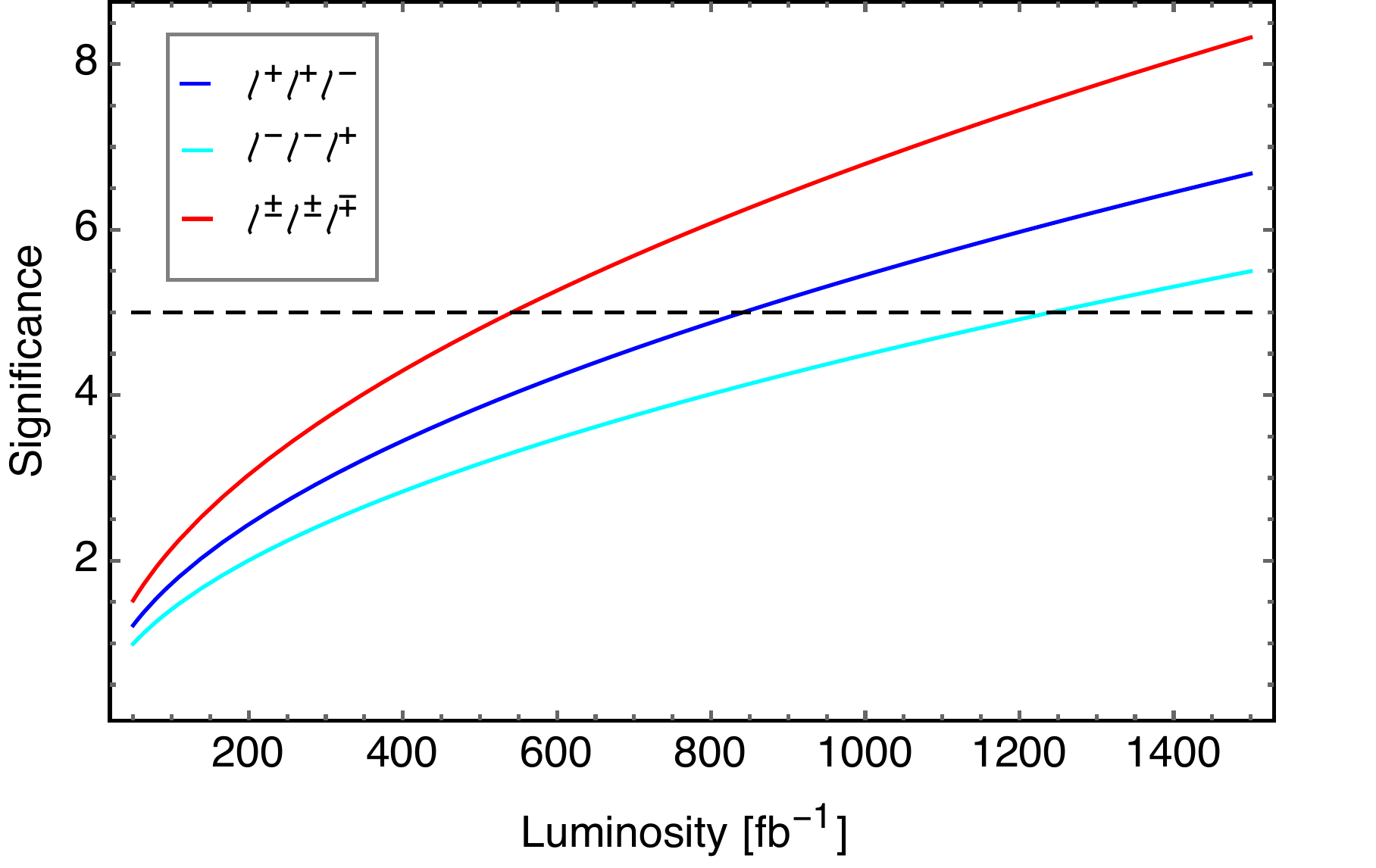}
  \hspace*{.7cm}\includegraphics[width=0.46\columnwidth]{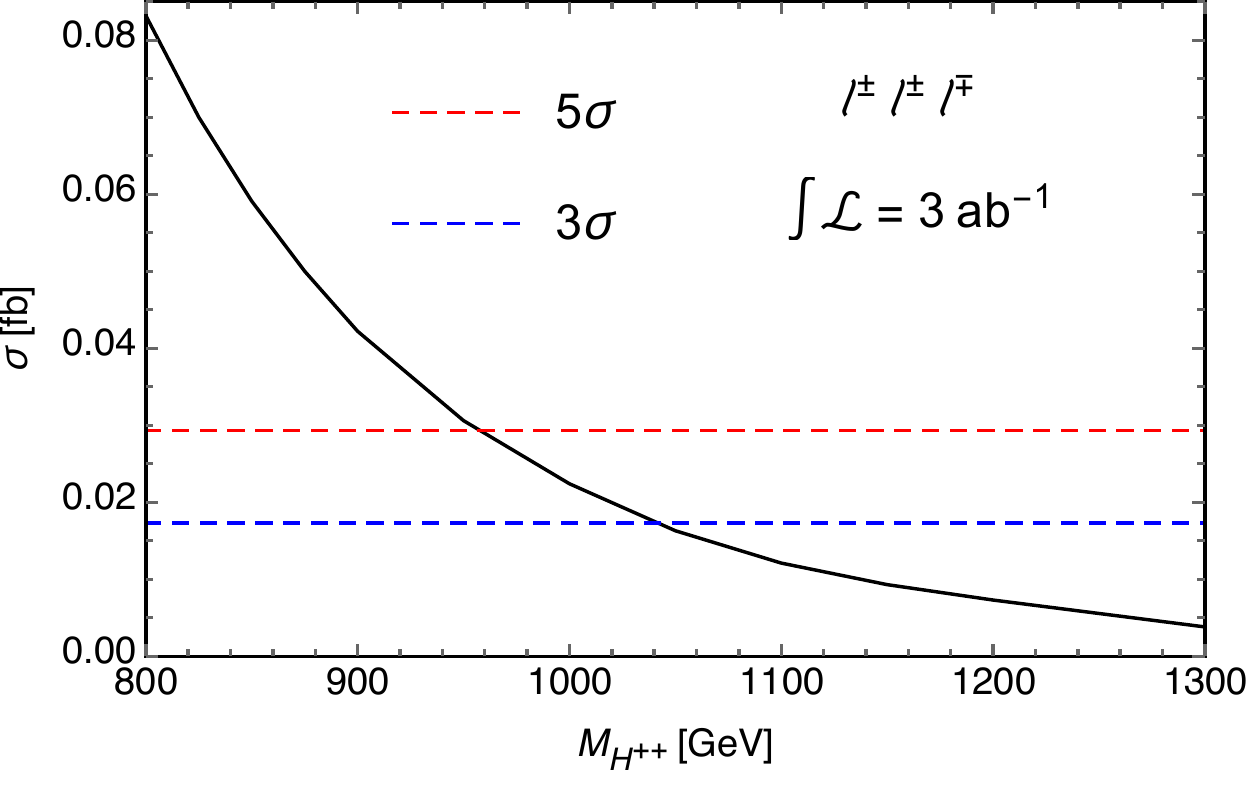}
  \caption{Left: {\it Dependence of the LHC sensitivity to the trileptonc MLRSM
    signal on the integrated luminosity. We separately indicate results for the
    $\ell^+ \ell^+\ell^-$ (blue) and $\ell^-\ell^-\ell^+$ (cyan) channels, as
    well as for their combination (red).} Right: {\it Dependence of the signal
    fiducial cross section after all selections on the double-charged Higgs
    boson mass $M_{H^{++}}$. The corresponding $3\sigma$ (blue) and $5\sigma$
    (red) reference lines are indicated, assuming an integrated luminosity of
    3~ab$^{-1}$.}}
  \label{fig_3l-obs}
\end{figure}

These selections are sufficient to get a decent sensitivity to the signal, as
shown in the left panel of Figure~\ref{fig_3l-obs} in which we present the
dependence of the significance $s$ calculated as in Eq.~\eqref{eq:sign} on the
integrated
luminosity. Both the $\ell^+ \ell^+\ell^-$ and $\ell^-\ell^-\ell^+$ channels are
expected to yield promising results, a $5\sigma$ discovery being reachable within
about $800-1200$~fb$^{-1}$ of proton-proton collisions at $\sqrt{s}=14$~TeV in both
cases. Combining the two channels, a signal may even be observed during the
earlier phase of the LHC Run~3, with a luminosity of about 500~fb$^{-1}$. While
promising, the trilepton channel is not as competitive as the four-lepton one.
Nevertheless, the option of a combination is conceivable and could potentially
lead to an even better expectation. Our results are generalised in the heavier
mass scale case on the right panel of Figure~\ref{fig_3l-obs}, in which we
present the dependence of the trilepton fiducial cross section (including thus
the selection efficiency) on the mass of the doubly-charged Higgs boson. In
contrast to the four-leptonic channel, the entire luminosity expected to be
collected during the high-luminosity run of the LHC will only allow us to
barely reach the TeV mass regime.

\section{Summary and Conclusions}
\label{conclusions}

Left-right symmetric models offer natural explanations for parity violation at
the electroweak scale and an elegant way to address neutrino masses through the
embedding of a seesaw mechanism. As a consequence of their symmetry breaking
pattern, they feature additional scalar fields including $SU(2)_L$ and $SU(2)_R$
triplets. Such an enriched scalar sector offers
various handles for discovering left-right symmetric new physics and to
distinguish it from other extensions of the Standard Model. In this work,
we have focused on the multileptonic collider signatures of this scalar sector
and estimated how the LHC could be sensitive to it in the upcoming years. By
carefully designing first a benchmark configuration viable relatively to
present data, we have found that the production of four-lepton and trilepton
systems is enhanced and could be used the main discovery mode of the
model, even if current constraints already push the new physics masses to a high
scale.

For our study, we have considered a scenario with doubly-charged scalar masses
fixed to 800 GeV, heavy right-handed neutrinos with masses of 12~TeV and a $W_R$
boson of 10~TeV. With all other model parameters set to guarantee consistency
with flavour data, we obtain a scenario featuring one light singly-charged
and two light neutral scalar bosons with large triplet components. Although the
cross sections associated with the production of these light doubly-charged and
singly-charged states have been found to be of about 0.1--1~fb, we
have shown, by relying on state-of-the-art Monte Carlo simulations, that a
simple selection strategy could allow for observing the resulting four-leptonic
and trileptonic signals within the reach of the high-luminosity phase of the
LHC. In other words, MLRSM singly-charged and
doubly-charged Higgs bosons lying in the  TeV range could be reached in a
not too far future thanks to an analysis strategy yielding an
almost background-free environment.

\medskip

\acknowledgments

PP and DG acknowledge support from the DST-SERB, India project grant
EMR/2015/000333, and from the DST-FIST grant SR/FST/PSIl-020/2009 for
offering the computing resources needed by this work. BF has been partly
supported by French state funds managed by the Agence Nationale de la Recherche
(ANR) in the context of the LABEX ILP (ANR-11-IDEX-0004-02, ANR-10-LABX-63).

\bibliography{biblio}

\end{document}